\documentclass[a4paper,11pt]{article}
\usepackage{pos}
\usepackage{wrapfig}
\usepackage{graphicx}
\usepackage{subcaption}
\usepackage{aas_macros}
\usepackage{xcolor}

\title{Cosmic Rays in Galaxy Halos: 
Impacts on Galactic Outflows and Baryon Cycling}
\ShortTitle{Cosmic Rays in Galaxy Halos}

\author*[a,b]{Ellis R. Owen}
\author[c,d,e]{Leonard E. C. Romano}
\author[b,f,g,h,i]{Kentaro Nagamine}

\affiliation[a]{Astrophysical Big Bang Laboratory (ABBL), RIKEN Pioneering Research Institute (PRI), \\ Wak\={o}, Saitama, 351-0198 Japan}

\affiliation[b]{Theoretical Astrophysics, Department of Earth and Space Science, The University of Osaka, 1-1 Machikaneyama, Toyonaka, Osaka 560-0043, Japan}

\affiliation[c]{Universitäts-Sternwarte, Fakultät für Physik, Ludwig-Maximilians-Universität München, Scheinerstr. 1, D-81679 München, Germany}

\affiliation[d]{Max-Planck-Institut für extraterrestrische Physik, Giessenbachstr. 1, D-85741 Garching, Germany}

\affiliation[e]{Excellence Cluster ORIGINS, Boltzmannstr. 2, D-85748 Garching, Germany}

\affiliation[f]{Theoretical Joint Research, Forefront Research Center, The University of Osaka, 1-1 Machikaneyama, Toyonaka, Osaka 560-0043, Japan}

\affiliation[g]{Kavli Institute for the Physics and Mathematics of the Universe (WPI), UTIAS, \\ The University of Tokyo, Kashiwa, Chiba 277-8583, Japan}

\affiliation[h]{Department of Physics and Astronomy, University of Nevada, Las Vegas, 4505 S. Maryland Pkwy, Las Vegas, NV 89154-4002, USA}

\affiliation[i]{Nevada Center for Astrophysics, 
            University of Nevada, Las Vegas, 4505 S. Maryland Pkwy, Las Vegas, NV 89154-4002, USA}

\emailAdd{ellis.owen@riken.jp}

\abstract{Galaxies with high star-formation rate surface densities often host large-scale outflows that redistribute energy, momentum, and baryons between the interstellar medium and the halo, making them a key feedback channel regulating galaxy evolution. Despite their importance, the driving physics behind galactic outflows and their interaction with the surrounding halo is yet to be fully understood. In particular, the influence of a pre-existing reservoir of cosmic rays (CRs) in galaxy halos has not been clearly established. We determine the conditions required to launch outflows in the presence of halo CRs and investigate how CR pressure gradients modify outflow speeds. We find that CR halos suppress the development of large-scale, CR-driven winds and redirect CR feedback toward local recycling flows. Slow outflows are therefore more likely in young galaxies lacking extended CR halos, while fast winds in intense starbursts are dominated by momentum injection and largely unaffected by halo CRs.}

\ConferenceLogo{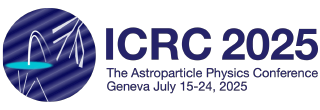}

\FullConference{39th International Cosmic Ray Conference (ICRC2025)\\
 15–24 July 2025\\
Geneva, Switzerland\\}
\begin{document}
\maketitle

\section{Introduction}
\label{sec:intro}

\noindent
Feedback from concentrated episodes of star formation can drive large-scale galactic winds. Such outflow winds are observed in nearby starbursts (e.g. M82) and are ubiquitous at high redshift, where galaxies are more compact and have higher star-formation rates. Outflow physical properties vary widely: flow speeds can reach up to $\sim$1,000 km s$^{-1}$, and densities/mass-loading factors vary by orders of magnitude between systems \cite{2015ApJ...809..147H, 2023ApJ...948...28X}. This diversity reflects variations in the energy, momentum, and matter supplied to the wind \cite{2018Galax...6..114Z, 2024ARA&A..62..529T}, differences in the flow-driving microphysics \cite{2021MNRAS.508.5092Y}, and the conditions of the surrounding galaxy halo. In this work, we focus on the effect of the halo on outflow development.

Hot halo gas exerts an inward thermal pressure.  This can oppose the development of an outflow by reducing its velocity and limiting its extension compared to systems without a halo~\cite{2021ApJ...917...12S}. By confining metal-enriched ejecta and restricting their dispersal, ram pressure from thermal gas in the halo has been considered to have an important role in baryon cycling and enriching circum-galactic medium (CGM) gas \cite{2005ApJ...634L..37F}.

\begin{figure}
\centering
 \includegraphics[width=0.6\linewidth]{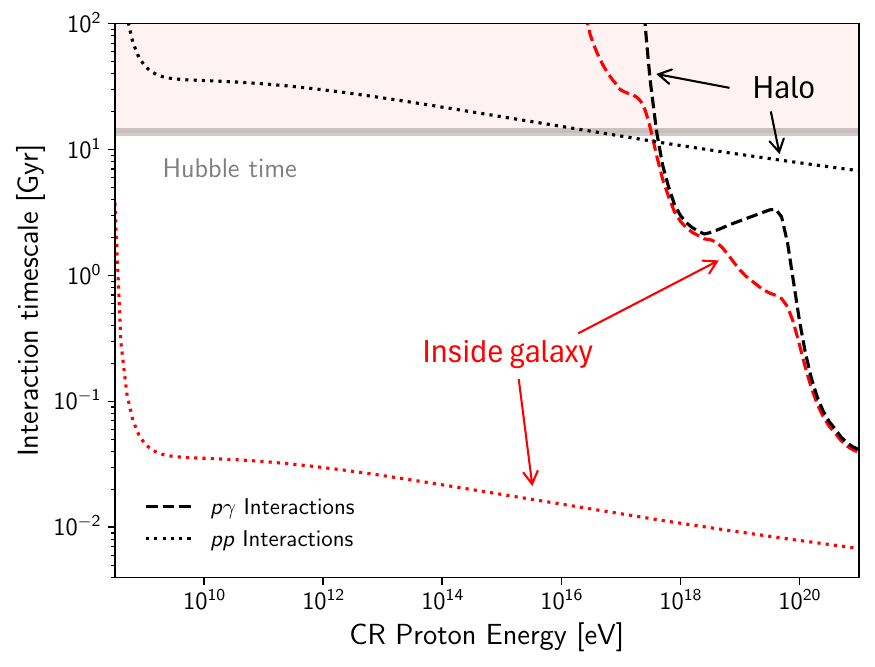}
 \caption{Characteristic interaction timescales for CR protons undergoing hadronic (pp) or photohadronic (p$\gamma$) losses under typical conditions in the galaxy interior (red) and halo (black). Interior conditions adopt a gas number density of
$n=1$ cm$^{-3}$ 
 and stellar radiation fields with a 7100 K stellar component ($0.7$ eV cm$^{-3}$) and 
 a 60 K dust component ($0.3$ eV cm$^{-3}$). Halo conditions adopt $n=10^{-3}$ cm$^{-3}$ 
and radiation fields reduced by a factor of 100 compared to the galaxy interior. p$\gamma$ losses on the CMB at $z=0$ are included in both cases. The Hubble time is shown in gray; interaction timescales longer than this (pink region) are effectively negligible. For a full discussion of the hadronic processes and timescale calculations, see \cite{Owen2018MNRAS}. Figure from \cite{Romano2025A&A}.} 
\label{fig:survival}
\end{figure}

In addition to thermal pressure, galaxy halos have been considered to host a reservoir of cosmic rays (CRs) in their CGM~\cite[e.g.][]{2021ApJ...914..135R, 2024arXiv241002066P, Roy2022ApJ, Ponnada2025arXiv250902697P}. These CRs may be transported from the host galaxy by advection in winds, buoyant bubbles, or AGN-driven activity \cite[e.g.][]{2019MNRAS.484.1645O, 2021ApJ...914..135R, 2022ApJ...926....8S, Ponnada2025arXiv250902697P}. Hadrons carry the bulk of this CR energy density, and have long lifetimes against absorption losses in the low-density halo (Fig. \ref{fig:survival}). They also only experience modest adiabatic, streaming and diffusive energy losses in halo conditions \cite[e.g.][]{Chan2019MNRAS}, meaning that they can establish a large-scale, non-thermal pressure gradient that acts with the hot halo gas to frustrate the eruption of galactic winds, or may render CR pressure gradients ineffective in driving a wind (this may account for studies that show nearby starburst winds, e.g. M82, do not require a CR driving component, even though CRs are likely to be abundant in these environments~\cite[e.g.][]{Wang2024arXiv241209452W}).

\section{Model}
\label{sec:model}

\subsection{Outflow development}
\label{sec:outflow_development}

\noindent
Feedback from a central starburst first drives a blastwave. If energy injection persists, this can develop into a sustained galactic wind.  To model this, we use the blastwave equation of motion \cite{1988RvMP...60....1O} for an outflow erupting from a starburst nucleus into a galactic halo:
\begin{equation}
{\underbrace{M \frac{d v_s}{dt}}_{\text{Acceleration}}} =
{\underbrace{F_{\rm SB} - \dot{M}_{\rm SB} v_s}_{\text{Wind Source}}} +
{\underbrace{\Delta P_{\rm CR} z_s^2}_{\text{Cosmic Rays}}} +
{\underbrace{\Delta P_s z_s^2 - \rho_0 z_s^2 (v_s - v_0)^2}_{\text{Gas Pressure \& Drag}}} -
{\underbrace{M g}_{\text{Gravity}}} \ .
\end{equation}
Here, $v_{\rm s}$ and $z_{\rm s}$ are the instantaneous velocity and vertical position of the swept-up blastwave shell, $M$ is its mass, and $F_{\rm SB}$ is the effective driving force supplied by the starburst winds. The terms $\Delta P_{\rm CR}$ and $\Delta P_{\rm s}$ are the CR and thermal pressure differences between the shocked and un-shocked gas. The drag term accounts for the ram pressure of ambient halo gas of density $\rho_0$, with $v_0$ denoting its initial velocity (for our purposes, we consider this is negligible). The final term describes the effect of the  
galaxy's gravitational field. 

\begin{figure}
\begin{center}
 \includegraphics[width=0.9\linewidth, clip=true]{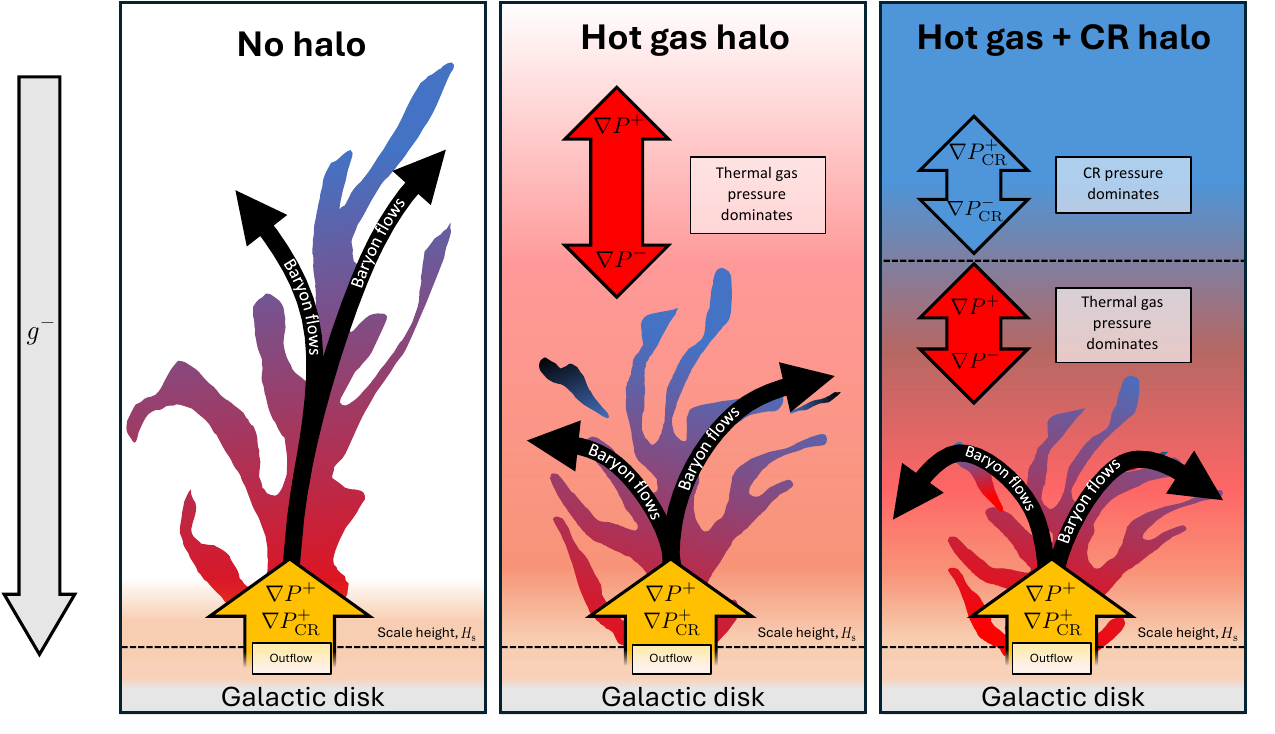}
 \end{center}
 \caption{Schematic of a starburst-driven outflow interacting with its galactic halo. The scale height and the opposing gravitational field 
$g$ are indicated. Superscripts ``+'' and ``–'' mark terms that drive or hinder the flow, respectively.
\textbf{Left}: In the absence of a substantial halo, the outflow expands freely and can escape the galaxy’s potential.
\textbf{Center}: As stellar mass builds, feedback produces a hot gas halo that confines the wind, enhancing baryon recycling and enriching the CGM \citep{2005ApJ...634L..37F, 2021ApJ...917...12S}.
\textbf{Right}: If CRs are present in the halo, they accumulate and establish a non-thermal pressure component with a longer characteristic scale height than the thermal gas. The outflow then encounters two distinct confining layers: thermal and CR-dominated. This further reduces its capability to escape. Figure from Ref. \cite{Romano2025A&A}.} 
 \label{fig:schematic_flows}
 \vspace{-0.5cm}
\end{figure}

We evaluate the steady-state configuration of the flow, explicitly accounting for the influence of the hot halo gas and CR halo (see Fig. \ref{fig:schematic_flows}). 
The starburst supplies mass and energy to the shell at rates of $\dot{M}_{\text{SB}} = M_{\text{ej}} \mathcal{R}_{\text{SN}}$, $\dot{E}_{\text{SB}} = E_{\text{SN}} \mathcal{R}_{\text{SN}}$, where $M_{\text{ej}} = M_{\text{ej, 0}} \, \text{M}_{\odot}$ is the typical supernova ejecta mass and $E_{\text{SN}} = 10^{51}\, E_{51} \, \text{erg}$ is the mechanical energy per supernova. These can be related to the star-formation rate (SFR) via the mass- and energy-loading factors, $\eta_{\text{m}} \equiv \dot{M}_{\text{SB}} / \mathcal{R}_{\text{SF}} = 0.01 M_{\text{ej, 0}}$ and $\eta_{\text{e}} \equiv \varepsilon_{\text{w}} \dot{E}_{\text{SB}} / \left(E_{\text{SN}} \mathcal{R}_{\text{SN}}\right) = \varepsilon_{\text{w}}$, respectively, where $\varepsilon_{w}$ is a thermalization efficiency factor accounting for energy dissipation in the system \citep[see e.g.][]{2016MNRAS.455.1830T, 2017ApJ...834...25K, 2024ApJ...960..100S}. For convenience we introduce scaled quantities $\eta_{\text{m, -2}} =\eta_{\text{m}}/0.01$,  $\eta_{\text{e, -2}} = \eta_{\text{e}} / 0.01$ and $\varepsilon_{\text{w, -2}} = \varepsilon_{\text{w}} / 0.01$. The fraction of energy injected into CRs is parametrized by $f_{\rm{CR}}$, defined at the galactic mid-plane. In our model, $\eta_{\rm m}$, $\eta_{\rm e}$, $\varepsilon_{\rm w}$ and $f_{\rm CR}$ are treated as constants during the outflow evolution, such that the dynamics are governed by the combined effects of thermal pressure, kinetic energy injection, and CR pressure gradients.

\vspace{-0.2cm}
\subsection{Cosmic Ray Halo Model and Transport}
\label{sec:results}

\noindent
CR pressure most strongly influences outflow development 
at low altitudes. To achieve a sensible normalization in this region, we set the halo CR content to match the mid-plane galactic CR energy fraction. The resulting halo CR pressure is then $P_{\rm CR,ext} = \frac{\gamma_{\rm CR}-1}{\gamma_{\rm th}-1}\, f_{\rm CR}\,\rho_{\rm mp}\,\sigma^{2}$, where $\gamma_{\rm th}$ and $\gamma_{\rm CR}$ are the adiabatic indices of the thermal and CR components, $\rho_{\rm mp}$ is the mid-plane gas density, and $\sigma$ is the halo gas velocity dispersion. This prescription yields a layered halo structure: thermal pressure dominates at low altitudes, while CR pressure becomes increasingly important at larger distances from the galaxy (see Fig.~\ref{fig:schematic_flows}). 

\subsubsection{Transport Pathways}
\label{sec:transport_pathways}

\noindent
The transport of CRs from the galactic mid-plane to the
halo must proceed in such a way to prevent a large fraction of particles from returning 
to the disk, and produce a substantial CR filling factor. 

\vspace{0.2cm}
\noindent
\textbf{Advection} in galactic outflows can partially achieve this over timescales that are comparable to or less than the starburst episodes driving an outflow. For an outflow velocity of $v_{\rm out}\sim100~{\rm km\,s^{-1}}$, CRs can reach a height of $H=10$ kpc in $t_{\rm adv}\sim H/v_{\rm out}\sim 100$ Myr. AGN feedback channels could act on even shorter timescales \cite[e.g.][]{Ponnada2025arXiv250902697P}, though are not guaranteed to be present in all starburst systems. Moreover, achieving a large volume-filling fraction required to spread the CRs throughout the halo (up to heights of 100 kpc) would take much longer and presumably relies on CR diffusion in regions far beyond the outflow cone. 

\vspace{0.2cm}
\noindent
\textbf{Diffusion} may operate over similar timescales. For a typical GeV CR diffusion coefficient of $D\sim3\times10^{28}\ {\rm cm^{2}\,s^{-1}}$, the diffusion timescale is
$t_{\rm diff}\sim {H^{2}}/{4D}\sim 200$ Myr. This is 
comparable to the advection timescale, but yields a declining CR profile with altitude. In magnetized regions near the disk, $D$ may be lower, potentially making it difficult to maintain a large CR filling factor during starburst lifetimes ($\sim 100$ Myr \cite{DiMatteo2008A&A}) unless super-diffusion or streaming operates. 

\subsubsection{Buoyant Bubbles as CR Transporters}
\label{sec:buoyant_bubbles}

\noindent
A more efficient mechanism for building a volume-filling CR halo is the rise of CR-rich buoyant bubbles. These could be inflated by stellar winds and SN-driven activity  \cite[see][]{Strickland1999MNRAS, Nguyen2022ApJ, 2025PASJHerenz}. For a bubble in pressure equilibrium with the ambient medium, its terminal rise speed follows from buoyancy-drag balance:
\begin{equation}
v_{\rm bub}^{\rm th}=\sqrt{\frac{g\,R_{\rm bub}^{\rm th}(1-\eta_{\rm th})}{C_d}}
\end{equation}
where $R_{\rm bub}^{\rm th}$ is the bubble radius, $\eta_{\rm th}=\rho_{\rm bub}^{\rm th}/\rho_0$ is the under-density, and $C_d\sim1$ is a drag coefficient accounting for entrainment and shape distortions. 

If the bubble is CR-dominated, its density contrast declines more slowly with height and its volume grows slightly faster (adiabatic scaling $V\propto P_{\rm ext}^{-1/\gamma}$), maintaining buoyancy longer. Compared to a thermally-dominated bubble, its terminal velocity is increased to:  
\begin{equation}
v_{\rm bub}^{\rm CR}=v_{\rm bub}^{\rm th} \left(\frac{R_{\rm bub}^{\rm CR}}{R_{\rm bub}^{\rm th}}\right)^{1/2} \left(\frac{1-\eta_{\rm CR}}{1-\eta_{\rm th}}\right)^{1/2} \ .
\end{equation}
This implies CR bubbles maintain their buoyancy better, penetrate deeper into the halo and are less likely to stall than purely thermal bubbles.

As the bubbles rise, CRs escape into the halo through their walls by diffusion. The walls have a characteristic thickness $\ell_{\rm bub}\sim100$ pc (c.f. with magnetic-draping layers that confine CRs \cite{Dursi2008ApJ}). For $D_{\perp}\sim10^{28}\ {\rm cm^{2}\,s^{-1}}$ as the diffusion coefficient across the bubble's surface, the leakage timescale is $t_{\rm leak}\lesssim 1$ Myr. This is much shorter than the rise time, so CRs are deposited continuously along the bubble trajectory as it rises through the halo. For kpc-sized bubbles in a Milky Way–like halo, the rise time to $>10$ kpc is $\sim 100$ Myr, with CR-dominated bubbles rising about 5\% faster. This makes buoyant bubbles a viable mechanism for populating the a large volume of the inner halo up to 10s of kpc with CRs during $>100$ Myr starburst episodes. 

\section{Results}
\label{sec:results}

\begin{figure}
\centering
 \includegraphics[width=0.5\linewidth, clip=true]{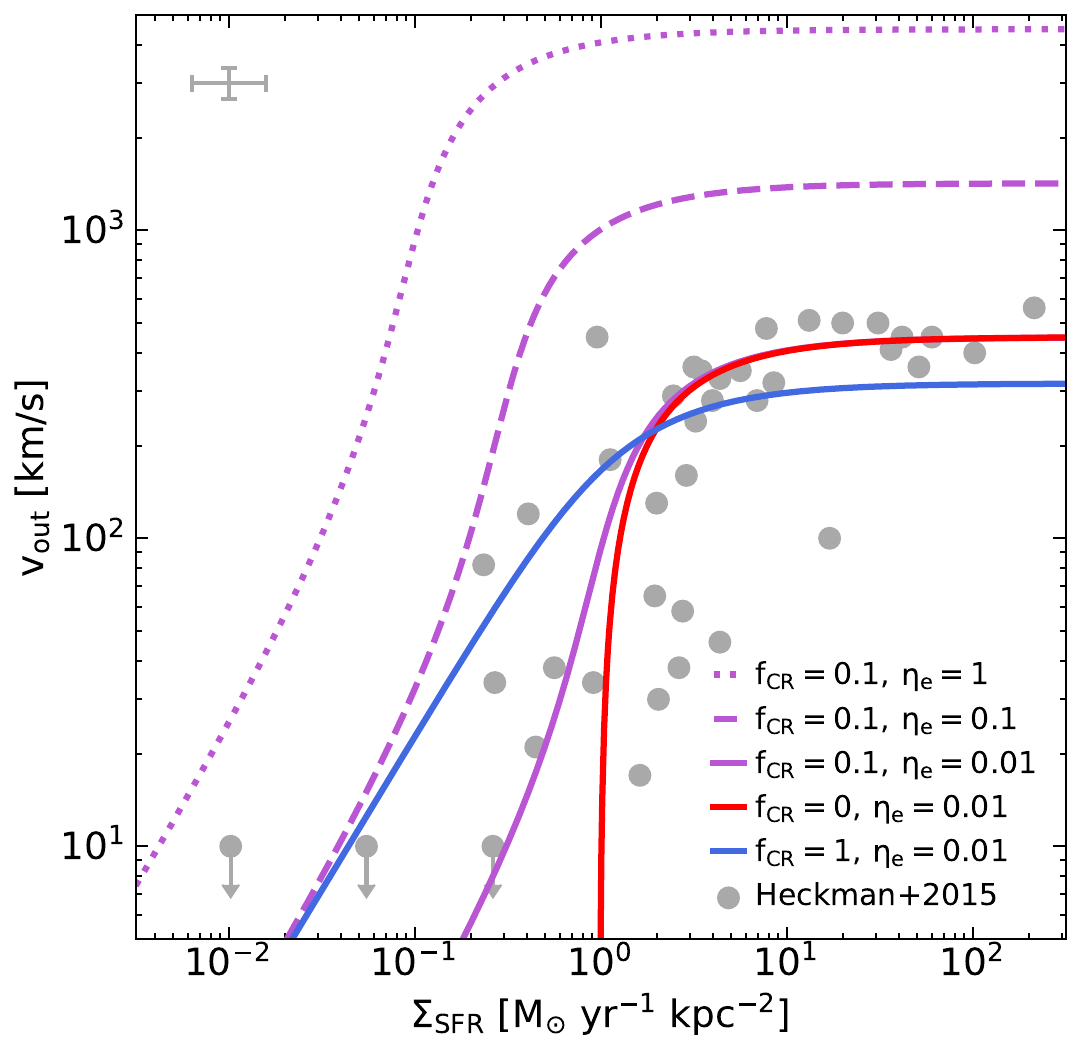}
 \caption{Predicted terminal outflow velocities as a function of star-formation rate surface density $\sum_{\rm SFR}$, for different CR energy fractions $f_{\rm CR}$
 and energy-loading factors $\eta_{\rm e}$. Observational data from \cite{2015ApJ...809..147H} are shown for a sample of nearby starburst galaxies with stellar masses 
$\text{log}_{10} \left( M_{*} / \text{M}_{\odot}\right) \in [ 7.1 - 10.9 ]$; typical uncertainties are indicated in the upper-left corner. Models with $\eta_{\rm e}\approx 0.01$ and $f_{\rm CR} \approx 0.1$ reproduce the observed trend, while higher $\eta_{\rm e}$ values 
 predict velocities that exceed the data. The transition from slow winds in weak starbursts to fast winds in strong starbursts is best captured by models including a non-zero CR fraction. In the absence of CRs, winds are launched only when the star-formation rate surface density exceeds a critical value of $\sim 1 \;\! {\rm M}_{\odot}\;\!{\rm yr}^{-1} \;\! {\rm kpc}^{-2}$, but slow winds can be sustained by CRs below this threshold. 
 } 
 \label{fig:CR_driven_outflows}
 \vspace{-0.3cm}
\end{figure}

\noindent
We solve the blastwave equation of motion for the parameterized mass, energy, and CR injection rates, assuming the system reaches a steady-state configuration. We apply the model to the galaxy sample of Ref. \cite{2015ApJ...809..147H} (Fig. \ref{fig:CR_driven_outflows}). Our key findings are: 
\begin{itemize}
    \item In the absence of CRs, winds are launched only when the star-formation rate surface density exceeds a critical value of $\sim 1 \;\! {\rm M}_{\odot}\;\!{\rm yr}^{-1} \;\! {\rm kpc}^{-2}$, which is set by the dynamic equilibrium pressure of the ISM-halo interface.
    \item CRs can sustain slow winds even below this critical threshold. We identify two distinct regimes: (i) Slow, CR-dominated outflows at subcritical SFR surface densities, and 
    (ii) Fast, momentum-driven outflows once the SFR exceeds the threshold. 
    \item The presence of an extended CR halo suppresses CR-driven winds beyond the galactic scale height. Instead, CRs refocus the influence of winds toward local recycling flows and galactic fountains, reducing the net mass and energy escape to the CGM. 
\end{itemize}

\section{Discussion and implications}
\label{sec:discussion_implications}
\vspace{-0.3cm}

\noindent
Numerical simulations of starburst-driven winds generally find that the inclusion of CRs improves their ability to drive warm, mass-loaded outflows \citep[e.g.][]{2016ApJ...816L..19G, 2022MNRAS.517..597C, 2024ApJ...964...99A}. Our results are consistent with this picture: in the CR-dominated regime, the outflow velocity is largely independent of mass loading, indicating that CR-driven winds can sustain substantial mass loading before being appreciably slowed.
However, apart from recent exceptions~\cite[e.g.][]{Ponnada2025arXiv250902697P} simulations do not usually include a pre-existing CR halo in their initial conditions. This may lead to an overestimate in the ability of CRs to accelerate material out of the galaxy. 

Although our approach is simplified (see Ref. \cite{Romano2025A&A} for a full discussion of its limitations), it provides useful insights into the qualitative behavior of halo CRs in suppressing large-scale winds and redirecting energy into local fountains and recycling flows. Future work is needed to properly test these predictions with high-resolution simulations that include CR halos as an initial condition, allowing their dynamical influence on wind launching, confinement, and CGM enrichment to be explored self-consistently.

\vspace{0.4cm}

\noindent
\footnotesize
\textbf{Acknowledgements:} E.R.O acknowledges support from the RIKEN Special Postdoctoral Researcher Program for junior scientists. This research was funded in part by the Deutsche Forschungsgemeinschaft (DFG, German Research Foundation) under Germany's Excellence Strategy – EXC 2094 – 390783311, and MEXT/JSPS KAKENHI grant numbers 20H00180, 22K21349, 24H00002, and 24H00241 (K.N.). K.N. acknowledges the support from the Kavli IPMU, the World Premier Research Centre Initiative (WPI), UTIAS, the University of Tokyo.  
\normalsize

\bibliographystyle{ICRC}
\bibliography{references}

\end{document}